\documentclass{article}

\usepackage{amssymb}

\usepackage{epsfig}

\usepackage{lscape}

\usepackage{graphicx,psfrag,amsmath}

\usepackage{yfonts}

\begin{document}

\def\beq#1\eeq{\begin{equation}#1\end{equation}}
\def\beql#1#2\eeql{\begin{equation}\label{#1}#2\end{equation}}

\def\bea#1\eea{\begin{eqnarray}#1\end{eqnarray}}
\def\beal#1#2\eeal{\begin{eqnarray}\label{#1}#2\end{eqnarray}}

\newcommand{\Z}{{\mathbb Z}}
\newcommand{\N}{{\mathbb N}}
\newcommand{\C}{{\mathbb C}}
\newcommand{\Cs}{{\mathbb C}^{*}}
\newcommand{\R}{{\mathbb R}}
\newcommand{\intT}{\int_{[-\pi,\pi]^2}dt_1dt_2}
\newcommand{\Cc}{{\mathcal C}}
\newcommand{\cI}{{\mathcal I}}
\newcommand{\cN}{{\mathcal N}}
\newcommand{\cE}{{\mathcal E}}
\newcommand{\Ca}{{\mathcal A}}
\newcommand{\xdT}{\dot{{\bf x}}^T}
\newcommand{\bDe}{{\bf \Delta}}

\def\ket#1{\left| #1\right\rangle }
\def\bra#1{\left\langle #1\right| }
\def\braket#1#2{\left\langle #1\vphantom{#2}
  \right. \kern-2.5pt\left| #2\vphantom{#1}\right\rangle }
\newcommand{\gme}[3]{\bra{#1}#3\ket{#2}}
\newcommand{\ome}[2]{\gme{#1}{#2}{\mathcal{O}}}
\newcommand{\spr}[2]{\braket{#1}{#2}}
\newcommand{\eq}[1]{Eq\,\ref{#1}}
\newcommand{\xp}[1]{e^{#1}}

\def\limfunc#1{\mathop{\rm #1}}
\def\Tr{\limfunc{Tr}}

\def\dr{detector }
\def\drn{detector}
\def\dtn{detection }
\def\dtnn{detection}

\def\pho{photon }
\def\phon{photon}
\def\phos{photons }
\def\phosn{photons}
\def\mmt{measurement }
\def\an{amplitude}
\def\a{amplitude }
\def\co{coherence }
\def\con{coherence}

\def\st{state }
\def\stn{state}
\def\sts{states }
\def\stsn{states}

\def\cow{"collapse of the wavefunction"}
\def\de{decoherence }
\def\den{decoherence}
\def\dm{density matrix }
\def\dmn{density matrix}

\newcommand{\mop}{\cal O }
\newcommand{\dt}{{d\over dt}}
\def\qm{quantum mechanics }
\def\qms{quantum mechanics }
\def\qml{quantum mechanical }

\def\qmn{quantum mechanics}
\def\mmtn{measurement}
\def\pow{preparation of the wavefunction }

\def\me{ L.~Stodolsky }
\def\T{temperature }
\def\Tn{temperature}
\def\t{time }
\def\tn{time}
\def\wfs{wavefunctions }
\def\wf{wavefunction }
\def\wfn{wavefunction} 
\def\wfsn{wavefunctions}
\def\wvp{wavepacket }
\def\pa{probability amplitude } 
\def\sy{system } 
\def\sys{systems }
\def\syn{system} 
\def\sysn{systems} 
\def\ha{hamiltonian }
\def\han{hamiltonian}
\def\rh{$\rho$ }
\def\rhn{$\rho$}
\def\op{$\cal O$ }
\def\opn{$\cal O$}
\def\yy{energy }
\def\yyn{energy}
\def\yys{energies }
\def\yysn{energies}
\def\pz{$\bf P$ }
\def\pzn{$\bf P$}
\def\pl{particle }
\def\pls{particles }
\def\pln{particle}
\def\plsn{particles}

\def\plz{polarization  }
\def\plzs{polarizations }
\def\plzn{polarization}
\def\plzsn{polarizations}

\def\sctg{scattering }
\def\sctgn{scattering}

\def\prob{probability }
\def\probn{probability}

\def\om{\omega} 

\def\hf{\tfrac{1}{2}}
\def\hft{\tiny \frac{1}{2}}

\def\zz{neutrino }
\def\zzn{neutrino}
\def\zzs{neutrinos }
\def\zzsn{neutrinos}

\def\zn{neutron }
\def\znn{neutron}
\def\zns{neutrons }
\def\znsn{neutrons}

\def\hf{\tfrac{1}{2}}

\def\csss{cross section }
\def\csssn{cross section}

\def\epp{elementary particle physics }
\def\eppn{elementary particle physics}

\def\vhe{very high energy }
\def\vhen{very high energy}

\def\bd{`black disc' }
\def\bdn{`black disc}

\title{ Highest Energy Proton-Nucleus\\ Cross Sections }

\author{L. Stodolsky,\\
Max-Planck-Institut f\"ur Physik
(Werner-Heisenberg-Institut)\\
F\"ohringer Ring 6, 80805 M\"unchen, Germany}

\maketitle

\begin{abstract}
 The description of \vhe proton-proton \csssn s in terms
of a \bd with an `edge' allows of a simple generalization to
highest energy proton-nucleus cross sections. This results in a
leading $ln^2W$ term and  a $ln\, W$ term whose coefficient depends
linearly on the radius of the nucleus ($W$ the c.m. \yy ). The
necessary parameters are determined from the fits to p-p data.
Since the coefficient of the $ln W$ term is rather large, it is
doubtful that the regime
of $ln^2W$ dominance can be reached with available \yys in
accelerators
or cosmic rays. However, the $ln W$ term can be relevant
for highest \yy cosmic rays in the atmosphere, where a large
increase for the \csss on nitrogen is expected. Tests of the theory
should be possible by
studying the coefficient  of $ln W$  at p-nucleus colliders.

\end{abstract}

\section{Introduction}

In recent years a simple picture for \vhe p-p \csssn s has been
developed,
in good agreement with experimental information. A short review is
presented
in Ref.\,\cite{beh}. It consists of a \bd \cite{comp} with a smooth
`edge' \cite{edge}. The
radius
of the `disc'  is growing as $ln\, W$ with
p-p  center-of-mass  \yy $W$, while the `edge' is
 constant with \yyn, with a thickness
 $t \approx 1.1 f$. The growing  `disc' contribution
has however a small coefficient,  with only a contribution 
$\sigma_{disc}^{TOT}\approx 1.1 mb \times
ln^2(W/m)$ to the total \csss \cite{comp}. Thus, although it
finally
dominates the more slowly growing
$\sim ln(W/m)$ `edge' contribution, it does not do so 
until the $W\approx 10\, TeV$ regime.

\begin{figure}[h]
\includegraphics[width=\linewidth]{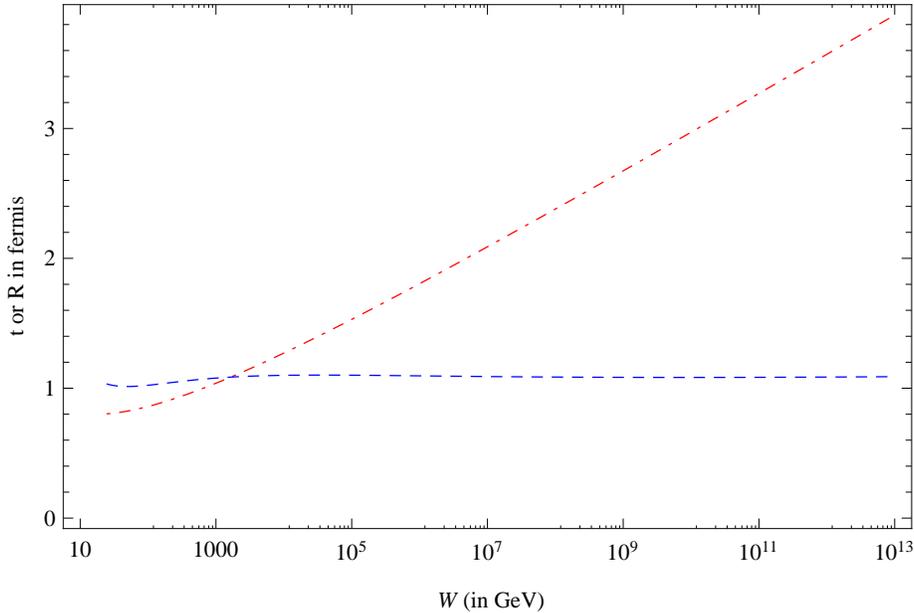}
\caption{The ``edge'' and the ``disc''. The dashed (blue) line is
a plot of the quantity $t$ from Ref.\,\cite{edge}, 
 representing, the effective thickness  of the edge.  Its
constancy  exhibits the 
\yy independence of the edge. 
The dashed-dotted  (red)  line represents  the
radius  inferred from the total \csssn, 
$\surd(\sigma^{TOT}/2\pi)$. The units are $f=fermi =10^{-
13}cm$.  From ref \cite{edge}, which used data from fits for the
total and 
elastic \csssn s. }

\label{ratio}
\end{figure}

This situation is shown in
Fig.\,\ref{ratio}, taken from \cite{edge}. The radius
$\surd(\sigma^{TOT}/2\pi)$ corresponding to the total \csss bceomes
larger than  the `edge' only around several TeV.

\section{Nuclei}
This simple geometric picture can be easily transfered  to the case
where, instead of another proton,  a nucleus
is the target for a \vhe proton.

At `high \yy' ($W\sim GeV's$) the nucleus will absorb any incident
proton
up to the ordinary, low \yyn, radius of the nucleus $R_A$, giving
a \bd of radius $R_A$.

This radius is typically parameterized in terms of the mass number
$A$ as \cite{bm}
\beql{an}
R_A\approx 1.2f A^{1/3}\, ,
\eeql
although the `1.2' factor may vary somewhat with the author or the
application.

 At `\vhe' ($W\gtrsim 10\,TeV$) the growing size of the
nucleon must be taken into account. For nucleons in the interior
of the nucleus the increase in the \csss
has little effect since the absorbtion is any case maximal,`black',
and cannot be sensibly increased. However it will affect 
interactions  with the nucleons on the rim or outer edge of the
nucleus, (seen in the plane transverse to the incoming proton). We
take this increase to be governed by the paramters determined from
the data on p-p \sctgn, as illustrated in Fig,\,\ref{ratio}. At
\yys
where there is `disc dominance' for the p-p interaction, one
therefore has an effective \bd radius for the interaction on the
nucleus $ R^{vhe}_A,$  which is
\beql{rad}
R^{vhe}_A\approx R_A+R^{vhe}_{pp}\,,
\eeql

For $R^{vhe}_{pp}$ we use the fit values \cite{comp} for the
p-p \bdn, namely $\sigma^{TOT}\approx 1.1
mb \,\,\times ln^2 (W/m)$   which implies
\beql{rp}
R^{vhe}_{pp}=\sqrt{1.1mb/2\pi}\times ln (W/m)\approx 0.13f\times ln
(W/m)\, 
\eeql
for the radius of the `disc' in p-p \sctgn,  the asymptotic slope
of the dashed-dotted  (red)  line of  Fig. \ref{ratio}.

\section{Consequences } 
With these parameters established, we examine
the consequences for nuclei. In  particular we consider the example
of nitrogen ($A=14$) as representative of cosmic ray interactions
in the atmosphere.

We look at the inelastic \csssn, $\sigma=\pi(R^{vhe}_A)^2$   as
relevant for  atmospheric interactions. (Adding the very forward
narrowly
peaked elastic \csss would give the total \csssn,
$2\pi(R^{vhe}_A)^2$.)

\beal{sig}
\sigma=\pi(R^{vhe}_A)^2 &=& const. +  0.26 \pi 
R_A\,f\, \times ln
(W/m)+\sigma^{disc}_{pp}\\ 
\nonumber
&=&const. +  0.26 \pi 
R_A\,f\, \times ln
(W/m)+ 0.55\, mb\,\,\times ln^2 (W/m).
\eeal

We thus arrive at an interesting  situation  with a small
$\ln^2(W/m) $ term and a  $ln(W/m)$ term with a relatively large
coefficient, namely $0.26\pi f R_A  $.

Evaluating this for nitrogen, $A=14$ using \eq{an}, one obtains
\beql{xterm}
 0.26  \pi \, R_A f \times ln(W/m)=  24\, mb\, \times ln(W/m)
\eeql

Although the  $\ln^2(W/m) $ term is ultimately dominant, it
will probably never be directly observable, due to its small
 coefficient.  With the value  \eq{xterm} the $\ln^2(W/m) $ term
in \eq{sig} becomes comparable to the $\ln(W/m)$ term  when \\ 
$ln(W/m)\approx
24/0.55= 43$, that is,
at $W\sim 10^{18} GeV$. This is much beyond the highest
\yy at the LHC or  the cosmic ray cutoff around $W\approx
\sqrt{10^{12}GeV^2}=10^6 GeV$.

It thus appears that for conceivably available \yys the
 \csssn s will be governed by the $ln (W/m)$ term, exhibiting 
effectively a
$const. + ln (W/m)$ behavior. Despite this only linear behavior in
the logarithm,  the relatively large
coefficient  induced by the nuclear radius means 
there can  nevertheless be significant effects, at \vhen. If we
inquire as to at
what \yy the growing \eq{xterm} becomes comparable to the low \yy
\csss $\pi R_A^2$, one has the condition
\beql{inc}
\frac{ \pi R_A 0.26 f\, \times ln(W/m)}{\pi R_A^2}= \frac{0.26
f}{R_A}\times ln(W/m)\sim 1\,,
\eeql
or when $ln(W/m)\sim \frac{R_A}{0.26 f}$. For
nitrogen this gives
 $W\sim 6.3\times 10^4 GeV\approx 60\, TeV,$ which is in the range
considered for some future accelerators \cite{cour}. For cosmic
rays this
corresponds to a lab \yy
 in the proton-proton \sy of $2.1 \times 10^{18}eV$,
 in  the region of highest
\yy cosmic rays, which extends up to $\sim 10^{21}eV$. Taken at
face value \eq{inc}
 implies a doubling of the simple \csss $\pi R_A^2$ in this
region.
Such an increase would lead to air showers starting higher in the
atmosphere and so resemble a change in the chemical composition of
the   cosmic rays \cite{auger}.

\section{Experimental Checks}
Experimental checks of this theory
 could possibly be
carried out  at a p-nucleus collider.

At  \yys
sufficiently
high so that the `disc' dominates the pp \csssn, that is, beginning
around  $W\sim 10\, TeV$,
one may envision checking the following points\\
 a) The \csssn s vary with \yy approximately as $const. +ln
(W/m)$\\
 b) When comparing different nuclei the coefficient of the
$ln(W/m)$ is linearly proportional to the radius of the
nucleus\\  

   These effects could be studied by direct  measurement
of the \csssn s or by examining the narrowing of the elastic
diffraction peak,  corresponding to the increasing radius.
Verification of these points would be a strong validation
of our simple geometrical picture and provide an important input
for analyzing air showers.

Finally, we note  an   interesting feature of these arguments.  The
leading $ln^2 $ term  from p-p \sctg is
 reproduced for nuclei in \eq{sig}, and with the same coefficient.
This reinforces the impression (see \cite{beh}) that the
coefficient of the $ln^2$ term is a fundamental parameter, one that
ought to be calculable from 
an underlying  theory of hadron physics.

\section{Acknowlegement}

These ideas were stimulated by discussions at the {\it Astropysics
+ MAGIC  }meeting, La Palma, June 2018. I would like to thank
Razmik Mirzoyan and
the other  organizers for their invitation.

\end{document}